\documentclass[aps,prd,onecolumn,groupedaddress,showpacs,nofootinbib,amssymb]{revtex4}
\usepackage[T1]{fontenc}
\usepackage[latin1]{inputenc}
\usepackage{graphicx}
\usepackage[english]{babel}
\usepackage{amsmath}
\usepackage{amssymb}
\usepackage{amsfonts}

\begin{document}
\def\pp{{\, \mid \hskip -1.5mm =}}
\def\cL{{\cal L}}
\def\be{\begin{equation}}
\def\ee{\end{equation}}
\def\bea{\begin{eqnarray}}
\def\eea{\end{eqnarray}}
\def\beq{\begin{eqnarray}}
\def\eeq{\end{eqnarray}}
\def\tr{{\rm tr}\, }
\def\nn{\nonumber \\}
\def\e{{\rm e}}

\title{\textbf{Classifying and avoiding singularities  in the alternative
gravity  dark energy models }}

\author{S. Capozziello$^1$, M. De Laurentis$^1$, S. Nojiri$^2$,
S.D. Odintsov$^3$\footnote{Also at Center of Theoretical Physics,
TSPU, Tomsk, Russia.}}

\affiliation{\it $^1$Dipartimento di Scienze Fisiche, Università
di Napoli {}`` Federico II'', INFN Sez. di Napoli, Compl. Univ. di
Monte S. Angelo, Edificio G, Via Cinthia, I-80126, Napoli, Italy\\
$^2$ Department of Physics, Nagoya University, Nagoya 464-8602, Japan\\
$^3$Institucio Catalana de Recerca i Estudis Avancats (ICREA) and
Institut de Ciencies de l Espai (IEEC-CSIC), Campus UAB, Facultat
de Ciencies, Torre C5-Par-2a pl, E-08193 Bellaterra (Barcelona),
Spain.}

\date{\today}

\begin{abstract}
The future  finite-time singularities emerging in  alternative
gravity  dark energy models are classified and  studied in Jordan
and Einstein frames. It is shown that such singularity may occur even in
flat spacetime for the specific choice of the effective potential. The
conditions
for the avoidance of finite-time singularities are presented and
discussed.
The problem is reduced to
the study of a scalar field evolving on an effective potential by
using the conformal transformations. Some viable modified gravity models
are
analyzed in detail and the way to cure singularity is considered
by introducing the higher-order curvature corrections. These results maybe
relevant for the resolution of the conjectured problem in the relativistic
star formation in such modified gravity where finite-time singularity is
 also manifested.
 \end{abstract}


\maketitle


\vspace{5.mm}
\section{Introduction}
\label{intro}
Several assumptions of the Cosmological Standard Model have been
ruled out by the advent of the so called {\it precision cosmology}
capable of probing physics at very large redshifts.
The old picture, based upon radiation and baryonic
matter, has to be revised.  Beside the introduction of dark
matter, needed to fit the astrophysical dynamics at galactic and
galaxy cluster scales (i.e. to explain clustered structures), a
new ingredient is requested  in order to explain the observed
accelerated behavior of the Hubble flow: the so called dark energy.
Essentially, data coming from the luminosity distance of Ia Type
Supernovae \cite{sneIa}, the deep and wide galaxy surveys
\cite{lss} and the anisotropy of Cosmic Microwave Background
\cite{cmbr} suggest that the so-called Cosmological Concordance
Model ($\Lambda$CDM) is spatially flat, dominated by cold dark
matter (CDM $\sim 25\,\%)$ and dark energy $(\Lambda$
$\sim 70\,\%)$. The first ingredient should be able to
explain the dynamics of clustered structures while the latter, in
the form of an ``effective" cosmological constant, should give
rise to the late-time accelerated expansion.

The cosmological constant is the the most relevant candidate to
interpret the  cosmic acceleration, but, in order to overcome its
intrinsic shortcomings associated with the energy scale,
several alternative proposals have been
suggested (see recent reviews\cite{padma2003,PR03,copeland}):
quintessence models,
where the cosmic acceleration is generated by means of a scalar
field, in a way similar to the early time inflation, acting at
large scales and recent epochs \cite{quintessence}; models based
on exotic fluids like the Chaplygin-gas \cite{chaplygin}, or
non-perfect fluids \cite{cresc}; phantom fields, based on scalar
fields with anomalous signature in the kinetic term
\cite{phantom}, higher dimensional theories (braneworld)
\cite{brane}. All of these models have the common feature to
introduce  new sources into the cosmological dynamics, but, from
an ``economic" point of view, it would be preferable to develop
scenarios consistent with observations without invoking extra
parameters or components non-testable (up to now)  at a
fundamental level.

Alternative theories of gravity, which extend in some way
General Relativity (GR), allows to pursue this different approach
(no further unknown sources) giving rise to suitable cosmological
models where a late-time accelerated expansion is naturally
realized.

The idea that Einstein gravity should be extended or corrected
at large scales (infrared limit) or at high energies (ultraviolet
limit) is suggested by several theoretical and observational
aspects. Quantum field theories in curved spacetime, as well as
the low-energy limit of string theory, both imply
semi-classical effective Lagrangians containing  higher-order
curvature invariants or scalar-tensor terms. In addition, GR has
been definitely tested only at Solar System scales while it may show several
shortcomings, if checked at higher energies or larger scales.
Besides, in the opinion of several authors, the Solar System
experiments are not so conclusive to state that the {\it only
reliable theory at these scales is GR}.

Of course modifying the gravitational action asks for several
fundamental challenges. These models can exhibit instabilities
\cite{instabilities-f(R)} or ghost\,-\,like behaviors
\cite{ghost-f(R)}, while, on the other side, they should be
matched with observations and experiments in the low
energy limit (in other words, Solar System tests and
Parameterized Post Newtonian (PPN) limit should reproduce the
results of GR appropriately).
Despite of all these issues, in the last years,
several interesting results have been achieved in the framework of
the so called modified gravity at cosmological, galactic and solar
system scales (see Refs.~\cite{odirev,GRGrev}
for review).

For example, there exist cosmological solutions that lead to the
accelerated expansion of the Universe
at late times in specific models of $f(R)$
gravity as is discovered in refs.\cite{f(R)-noi,f(R)-cosmo,NO}.
In some of realistic theories of this sort the problems indicated in
\cite{Dolgov} maybe overcomed\cite{odirev}.

There exist viable $f(R)$ models that can satisfy both background
cosmological constraints
and stability conditions \cite{Nojiri:2007as,Li,AT07,HS,Appleby,Tsuji,
Nojiri:2007cq} as well as
local tests.
Recently many works have been devoted to place constraints on
$f(R)$-models using the observations of Cosmic Microwave
Background anisotropies and galaxy
power spectrum \cite{bean-f(R),Faul}.

Besides, considering $f(R)$-gravity in the low energy limit, it is
possible to obtain corrected gravitational potentials capable of
explaining the flat rotation curves of spiral galaxies and galaxy
cluster haloes without considering huge amounts of dark matter
\cite{Nojiri:2007as,noi-mnras,salucci,nojiri,mendoza,Boe,salzano}
and, furthermore, this seems the only self-consistent way to
reproduce the universal rotation curve of spiral galaxies
\cite{salucci2}. On the other hand, several anomalies in Solar
System experiments could be framed and addressed in this picture
\cite{bertolami}.

However, a fundamental task which has to be faced for any
alternative gravity model is to classify singularities which could
emerge at finite time and propose the way to avoid it.  From  a physical
point of view, this point
is crucial in order to achieve viable and self-consistent models,
especially in the possible applications.

In this paper, we discuss the future singularities which can, in
principle, appear in  dark energy models  coming from alternative
gravity theories (higher-order or non-minimally coupled gravity).

In fact, when dark energy
models with the effective equation of state parameter close to $-1$
were added to the list of admissible cosmological theories  to explain the observed accelerated behavior,
due to to violation of all (or part) of energy conditions,  strange features  emerged in the future.
For instance, it is  well-known that phantom dark energy brings the
universe to finite-time {\it Big-Rip} singularity \cite{mcinnes,CKW}.
Moreover, the
effective quintessence dark energy cosmologies may end up in (softer)
finite-time singularity \cite{Nojiri2005sx,barrow}. For such effective
quintessence dark energy models only part of energy conditions does not work in the standard way.
Nevertheless, they show up the rip singularity behaviors which have been
classified in Ref.\cite{Nojiri2005sx}.

It is clear that, qualitatively, the same situation should occur also in
 modified gravity cosmologies \cite{odirev}. Indeed, it is quite well-known that some versions of
modified gravity (like $f(R)$) have an effective ideal fluid
description \cite{capozziello}. Hence, precisely the same singular
behavior should be typical for the (effective phantom/quintessence)
modified gravities in future too. Indeed, it was found some time ago
\cite{briscese} that modified gravity becomes invalid (complex theory) at
the point where mathematically-equivalent scalar-tensor dark energy theory
enters to the Big Rip singularity. Moreover, the effective phantom behavior may
enter a transient phase and future singularity does not occur if some higher
order terms (like $R^2$) are added to initially phantom-like models
\cite{Abdalla}. The same approach has been recently considered in
ref.\cite{KobayashiMaeda} to remove the singularity in order to avoid the
conjectured problems with neutron stars formation in modified gravity.

In this paper, we want to discuss, in general, the problem  of
finite-time singularities and discuss some ways to avoid them in
viable models which well-fit data at local and cosmological
scales.

The layout of the paper is the following. In  Sect.\ref{2}, we
describe, in general, the problem of finite-time singularities in
dark energy models coming from $f(R)$ to scalar-tensor modified
gravities. Sect. \ref{3} is devoted to the  conditions for
singularity avoidance in    $f(R)$-gravity and in its
scalar-tensor counterpart. In Sect. \ref{4}, we  discuss the
singularity problem in some physically viable $f(R)$-models
adopting a conformal transformation approach. This method allows,
in principle, to discriminate singularities by studying the
behavior of effective scalar field potential after dynamics has
been conformally reduced to the Einstein frame. In particular, we
study the effect of adding a correction term, proportional to
$R^n$ with $n\geq2$, to modify the structure of the potential at
large values of $R$  and  cure the singularity.  Discussion and
conclusions are drawn in Sect. \ref{5}.
\noindent
\section{Finite-time singularities in dark energy models: from $F(R)$-gravity to
scalar-tensor theory}
\label{2}

Let us start with a generic action of $F(R)$-gravity which is a straightforward extension of General Relativity:
\be
\label{fr1}
S_{F(R)}=\int d^4 x \sqrt{-g} \left\{\frac{F(R)}{2\kappa^2} + {\cal L}_m\right\}\ .
\ee
Here  $F(R)=R+f(R)$ is an appropriate function of the scalar curvature $R$ and ${\cal L}_m$
is the Lagrangian density of matter.
By the variation over the metric tensor $g_{\mu\nu}$, we obtain the fourth-order field
equations:
\be
\label{JGRG13}
\frac{1}{2}g_{\mu\nu} F(R) - R_{\mu\nu} F'(R) - g_{\mu\nu} \Box F'(R)
+ \nabla_\mu \nabla_\nu F'(R) = - \frac{\kappa^2}{2}T_{(m)\mu\nu}\ .
\ee
In Eqs.(\ref{JGRG13}), $T_{(m)\mu\nu}$ is the matter energy-momentum
tensor. Contracting Eqs. (\ref{JGRG13}) with respect to $\mu$ and $\nu$, we
obtain the trace equation:
\be
\label{frlv1}
2F(R) - R F'(R) - 3 \Box F'(R) = - \frac{\kappa^2}{2} T\ .
\ee
To recover, formally, General Relativity, the Eq.(\ref{frlv1}) can be rewritten as
\be
\label{frlv2}
R + 2f(R) - Rf'(R) - 3\Box f'(R) = - \frac{\kappa^2}{2} T\ .
\ee
In order to study how finite-time singularities emerge and can be discussed, let us consider, for the moment,  classes of models which are paradigmatic for our purposes.
For example, in ref.\cite{HS}, a model which easily passes local tests
and several cosmological bounds, has been proposed:
\be
\label{HS1}
f_{HS}(R)=-\frac{m^2 c_1 \left(R/m^2\right)^n}{c_2 \left(R/m^2\right)^n + 1}\ ,
\ee
or otherwise written as
\begin{equation}
f_{HS}(R)=-\lambda
R_{c}\frac{\left(\frac{R}{R_{c}}\right)^{n}}{\left(\frac{R}{R_{c}}\right)^{n}+1}
\label{eq:HS1}
\end{equation}
Here $m$ is a proper scale and $c_1$, $c_2$, $n$ and $\lambda$ are dimensionless positive
constants ($n$ is not restricted to be an integer) and  $R_{c}$ is positive constant.
When the curvature is sufficiently large at dark energy epoch, this model
can be approximated as follows
\be
\label{frlv3}
f(R) \sim - 2\Lambda + \frac{\alpha}{R^n}\ .
\ee

In case of (\ref{HS1}), one may identify
\be
\label{HS2}
2\Lambda = \frac{m^2 c_1}{c_2}\ ,\quad \alpha = \frac{m^{2m + 2} c_1}{c_2^2}\ .
\ee
Then Eq.(\ref{frlv2}) reduces to
\be
\label{frlv4}
R + 3\alpha \Box \left(R^{-n-1}\right) \sim 0\ .
\ee
Since  the large curvature regime is considered, the cosmological constant
term appears as
a next-to-leading order correction, compared with the first term in
(\ref{frlv4}), and we neglect it.
If we define $\chi$ as
\be
\label{frlv5}
\chi \equiv R^{-n - 1}\ ,
\ee
and the FRW metric with flat spatial part is chosen,
\be
\label{fs0} ds^2 = - dt^2 + a(t)^2 \sum_{i=1,2,3}
\left(dx^i\right)^2\ .
\ee
Eq.(\ref{frlv4}) has the following
form:
\be
\label{frlv6}
\ddot \chi + 3H \dot \chi =
\frac{1}{3\alpha}\chi^{- 1/(n+1)}\ .
\ee
Note $\chi=0$ corresponds
to the curvature singularity $R\to\infty$. Note that as other dark
energy models with EoS parameter around $-1$, the above
gravitational alternative for dark energy also has the singularity
as it will be explained below. The fact that such $F(R)$-gravity
may show the phantom-like behavior has been established in
ref.\cite{Abdalla}. It was demonstrated there that account of
$R^2$ (or similar nature term) makes  the phantom phase to be
transient and removes the singularity. In principle, the phantom
phase in $F(R)$-gravity may end up as Big Rip-like type
singularity\cite{mcinnes} as it was demonstrated in
ref.\cite{briscese}.

First we consider the classical equation of motion:
\be
\label{frlv7}
\ddot x = \frac{1}{3\alpha}x^{- 1/(n+1)}\ .
\ee
The difference between the cosmological equation (\ref{frlv6})
and the classical equation (\ref{frlv7}) is the second term depending on $H$.
This second term gives the only sub-leading contribution, which will
be shown in
the analysis from (\ref{frlv9}), where the $H$ dependence will be explicitly
included and it will be shown that the result from the classical analysis here
will be reproduced.
For Eq.(\ref{frlv7}), one gets an exact solution:
\be
\label{frlv8}
x= C \left(t_0 - t\right)^{2(n+1)/(n+2)}\ .
\ee
Here $C$ and $t_0$ are constants. Note $2>2(n+1)/(n+2)>1$.
Then $x$ vanishes in a finite time $t=t_0$, which corresponds to the curvature singularity
in (\ref{frlv7}).

\begin{figure}[htbp]
\centering
\includegraphics[width=160mm]{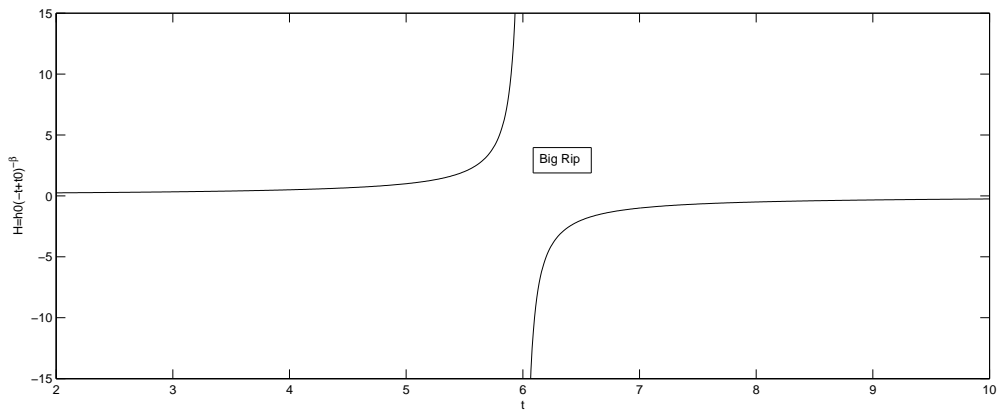}
\caption{Plot of  $H \sim h_0 \left(t_0 - t\right)^{-\beta}$. The
Big Rip singularity occurs for  $\beta=1$ and
$t=t_0=6$.}\label{fig:modelloeq14bigrip}
\end{figure}

We now investigate the asymptotic solution when the curvature is large, that is, $\chi$
is small. As there is a curvature singularity, one may assume
\be
\label{frlv9}
H \sim h_0 \left(t_0 - t\right)^{-\beta}\ .
\ee
Here $h_0$ and $\beta$ are constants, $h_0$ is assumed to be positive and
$t<t_0$ as it should be for the expanding universe.
Even for non-integer $\beta<0$, some derivative of
$H$ and therefore the curvature becomes singular. The case $\beta=1$
corresponds to the Big Rip singularity  Fig. \ref{fig:modelloeq14bigrip}.
Furthermore since $\beta=0$ corresponds to the de Sitter space, which has no singularity,
it is assumed $\beta\neq 0$.
When $\beta>1$, the scalar curvature $R$ behaves as
\be
\label{rlv16B}
R \sim 12 H^2 \sim 12h_0^2 \left( t_0 - t \right)^{-2\beta}\ .
\ee
On the other hand, when $\beta<1$, the scalar curvature $R$ behaves as
\be
\label{rlv16C}
R \sim 6\dot H \sim - 6h_0 \beta \left( t_0 - t \right)^{-\beta-1}\ .
\ee
Then $R$ diverges when $\beta>-1$ but $\beta\neq 0$.

\begin{figure}[htbp]
\centering
\includegraphics[width=160mm]{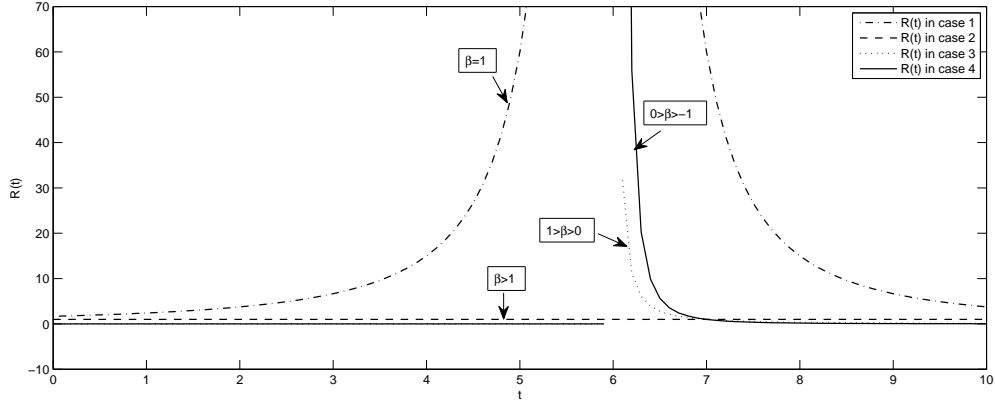}
\caption{Behavior of $R(t)$  for the four cases in the text. The
dash-dot line represents  $R$ for $\beta=1$ in Eq.(\ref{frl1});
the dashed line is for  $\beta>1 $ in  Eq.(\ref{frl5}); the dot
line is $R(t)$ for  $1>\beta>0$; finally,  the solid line $R(t)$
is for $0>\beta>-1$ in  Eq.(\ref{frl10})
respectively.}\label{fig:R}
\end{figure}

We now consider three cases (compare with \cite{nojiriprd}): 1) $\beta=1$,
2) $\beta>1$, 3) $1>\beta>0$,
and 4) $0>\beta>-1$ see Fig. \ref{fig:R}.
\begin{itemize}
\item In case 1) $\beta=1$, since
\be
\label{frl1}
R\sim \frac{12h_0^2 + 6h_0}{\left(t_0 - t\right)^{2}}\ ,
\ee
and therefore, from (\ref{frlv5}), we find
\be
\label{frl2}
\chi \sim \left(t_0 - t\right)^{2(n+1)}\ ,
\ee
and the l.h.s. of (\ref{frlv6}) behaves as
\be
\label{frl3}
\ddot \chi + 3H \dot \chi \sim \left(t_0 - t\right)^{2n}\ ,
\ee
but the r.h.s. behaves as
\be
\label{frl4}
\frac{1}{3\alpha}\chi^{- 1/(n+1)} \sim \left(t_0 - t\right)^{-2}\ ,
\ee
which is inconsistent since the powers of the both sides do not coincides
with each other. Therefore, $\beta\neq 1$.
\item In case 2) $\beta>1$, we find
\be
\label{frl5}
R = 12H^2 + 6\dot H \sim 12 H^2 \sim \left(t_0 - t\right)^{-2\beta}\ ,
\ee
and therefore
\be
\label{frl6}
\chi \sim \left(t_0 - t\right)^{2\beta(n+1)}\ .
\ee
In the l.h.s. of (\ref{frlv6}), the second term
dominates and the l.h.s. behaves as
\be
\label{frl7}
\ddot \chi + 3H \dot \chi \sim 3H \dot \chi \sim
\left(t_0 - t\right)^{\beta(2n+1) -1}\ .
\ee
On the other hand, the r.h.s. behaves as
\be
\label{frl8}
\frac{1}{3\alpha}\chi^{- 1/(n+1)} \sim \left(t_0 - t\right)^{-2\beta}\ .
\ee
Then by comparing the powers of the both sides, one gets
\be
\label{frl9}
\beta(2n+1) -1 = -2\beta\ ,
\ee
which gives $\beta = 1/(2n+3)$ but this conflicts with the assumption $\beta>1$.
\item In case 3) $1>\beta>0$ or case 4) $0>\beta>-1$, we find
\be
\label{frl10}
R = 12H^2 + 6\dot H \sim 6 \dot H \sim \left(t_0 - t\right)^{-\beta -1}\ ,
\ee
and therefore
\be
\label{frl11}
\chi \sim \left(t_0 - t\right)^{(\beta+1)(n+1)}\ .
\ee
Then in the l.h.s. of (\ref{frlv6}),
the first term dominates and the l.h.s. behaves as
\be
\label{frl12}
\ddot \chi + 3H \dot \chi \sim \ddot \chi
\sim \left(t_0 - t\right)^{\beta(n+1) + n -1}\ .
\ee
On the other hand, the r.h.s. behaves as
\be
\label{frl13}
\frac{1}{3\alpha}\chi^{- 1/(n+1)} \sim
\left(t_0 - t\right)^{-\beta -1}\ .
\ee
Then by comparing the powers of the left-hand side and the right-hand side,
the consistency gives
\be
\label{frl14}
\beta(n+1) + n - 1 = -\beta - 1\ \mbox{or}\ \beta = -n/(n+2)\ .
\ee
This conflicts with the case 3) $0<\beta<1$ but is
consistent with the case 4) $0>\beta>-1$.
In fact, by substituting (\ref{frl14}) into (\ref{frl11}), we get
\be
\label{frl14B}
\chi \sim \left(t_0 - t\right)^{2(n+1)/(n+2)}\ .
\ee
which corresponds to (\ref{frlv8}).
Since $0>\beta>-1$, this singularity corresponds to Type II in
\cite{Nojiri2005sx}.
\end{itemize}
Thus, the sudden finite-time curvature singularity really appears in the
Hu-Sawicki model.

In \cite{Nojiri2005sx}, there was suggested the classification of the
finite-time singularities in the following way:
\begin{itemize}
\item Type I (``Big Rip'') : For $t \to t_s$, $a \to \infty$,
$\rho \to \infty$ and $|p| \to \infty$. This also includes the case of
$\rho$, $p$ being finite at $t_s$.
\item Type II (``sudden'') : For $t \to t_s$, $a \to a_s$,
$\rho \to \rho_s$ and $|p| \to \infty$
\item Type III : For $t \to t_s$, $a \to a_s$,
$\rho \to \infty$ and $|p| \to \infty$
\item Type IV : For $t \to t_s$, $a \to a_s$,
$\rho \to 0$, $|p| \to 0$ and higher derivatives of $H$ diverge.
This also includes the case when $p$ ($\rho$) or both of them tend to
some finite values while higher derivatives of $H$ diverge.
\end{itemize}
Here $t_s$, $a_s$ and $\rho_s$ are constants with $a_s\neq 0$. We now identify $t_s$ with $t_0$.
The Type I corresponds to $\beta>1$ or $\beta=1$ case and we have a Big Rip singularity \cite{CKW}, whereas the Type II to $-1<\beta<0$ case and corresponds to the sudden future singularity, Type III
to $0<\beta<1$ case and is different from the sudden future singularity in the sense that $\rho$ diverges, and Type IV to $\beta<-1$ but $\beta$ is not any integer case.

\begin{figure}[htbp]
\centering
\includegraphics[width=180mm]{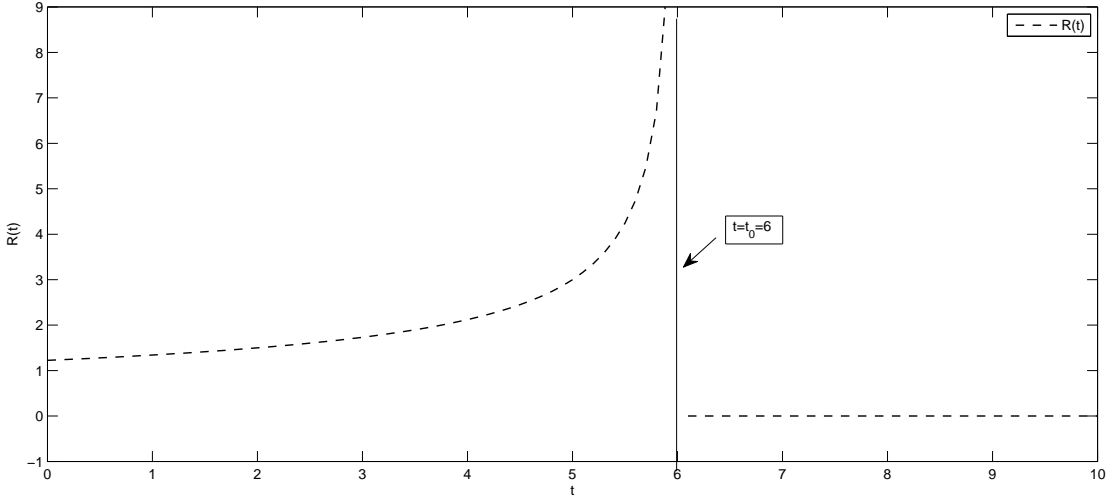}
\caption{Behavior of $R$ as given in Eq. (\ref{fs3}). We have
assumed $\gamma=\frac{1}{2}$.}\label{fig:modelloeq34}
\end{figure}

Let us now remind that the Type II singularity has been already discussed
in several dark energy models\cite{barrow} besides
$F(R)$-gravity.
Here we consider several theories in the FRW space-time with flat spatial
part (\ref{fs0}).
For the Type II  singularity
the Hubble rate $H \equiv \dot a /a$ has the following form
\be
\label{fs1}
H= H_0 + H_1 \left(t_0 - t\right)^\gamma \ .
\ee
Here $H_0$, $H_1$, $t_0$ and $\gamma$ are constants. We now choose $0<\gamma<1$. Then $H$ is finite
$H\to H_0$ in the limit of $t\to t_0$ but $\dot H$ diverges as
\be
\label{fs2}
\dot H = H_1 \gamma \left(t_0 - t\right)^{\gamma -1} \ ,
\ee
which generates the singularity in the scalar curvature $R$ since (see Fig.(\ref{fig:modelloeq34}))

\be
\label{fs3}
R=12 H^2 + 6 \dot H \sim 6 H_1 \gamma \left(t_0 - t\right)^{\gamma -1} \ .
\ee
We should note that the energy density $\rho$ is finite since the first FRW equation gives
\be
\label{fs4}
\rho = \frac{3}{\kappa^2}H^2\ ,
\ee
and therefore $\rho\to \left(3/\kappa^2\right) H_0^2 < \infty$ in the limit $t\to t_0$.
Hence, the curvature singularity could occur even if the energy density
is finite as in some other quintessence models.

We now give an explicit example of the ideal fluid which gives the
singularity in (\ref{fs1}). First we should note that the second
FRW equation has the following form: \be \label{fs5} p= -
\frac{1}{\kappa^2}\left(2\dot H + 3H^2\right)\ . \ee For the
Hubble rate $H$ in (\ref{fs1}), Eqs.(\ref{fs4}) and (\ref{fs5})
give \be \label{fs6} \rho = \frac{3}{\kappa^2}\left( H_0 + H_1
\left(t_0 - t\right)^\gamma\right)^2 \ ,\quad \rho + p = -
\frac{2H_1\gamma}{\kappa^2} \left(t_0 - t\right)^{\gamma -1} \ .
\ee Then by deleting $t$ in the two equations of (\ref{fs6}), we
find \be \label{fs7} \rho = \frac{3}{\kappa^2}\left( H_0 + H_1
\left( - \frac{\kappa^2\left(\rho + p\right)}{2H_1 \gamma}
\right)^{\gamma/\left(\gamma - 1\right)}\right)^2\ . \ee
Eq.(\ref{fs7}) can be regarded as an equation of state (EoS).
Conversely if we consider the perfect fluid satisfying the EoS
(\ref{fs7}), the singularity (\ref{fs1}) occurs.

We now consider the occurrence of singularity (\ref{fs1}), in terms of the
scalar-tensor theory, whose action is given by
\be
\label{fs8}
S=\int dx^{4}\sqrt{-g}\left[ \frac{1}{2\kappa^{2}}R
 - \frac{1}{2} \omega (\phi)
\partial_{\mu} \phi \partial^{\mu }\phi -V(\phi ) \right]\ ,
\ee
where $V(\phi )$ is the scalar potential
and $ \omega (\phi)$ is the kinetic function, respectively.
Note that for convenience the kinetic factor is introduced.
We should note that scalar field may be always redefined so
that kinetic function is absorbed.
In the spatially flat FRW space-time (\ref{fs0}), the energy density and the pressure
of the scalar field is given by
\be
\rho _{\phi} = \frac{1}{2} \omega (\phi )\, {\dot \phi}^{2}
+V(\phi)\ ,\qquad
p_{\phi } = \frac{1}{2} \omega (\phi ) \, {\dot \phi}^{2}-V(\phi)\ .
\label{eq:1.4}
\ee
Combining the FRW equations (\ref{fs4}) and (\ref{fs5}) with (\ref{eq:1.4}), one obtains
\be
\omega (\phi ) \, \dot{\phi ^{2}}
= -\frac{2}{\kappa^{2}}\dot{H} \ ,\qquad
V(\phi ) = \frac{1}{\kappa^{2}}
\left( {3H}^{2}+\dot{H} \right) \ .
\label{eq:1.5}
\ee
We now consider the theory in which $V(\phi)$ and $\omega(\phi)$ are given by
\be
\omega (\phi ) = -\frac{2}{\kappa ^{2}}f^{\prime }(\phi )\ ,\qquad
V(\phi ) = \frac{1}{\kappa ^{2}} \left[ {3f(\phi)}^{2}+f^{\prime }(\phi ) \right]\ ,
\label{eq:1.6a}
\ee
where $f(\phi)$ is a proper function of $\phi$. Then the following solution
is found
\be
\label{ST1}
\phi =t\ , \quad H(t)=f(t)\ .
\ee
In case of (\ref{fs1}), we find
\be
\label{fs9}
\omega (\phi) = - \frac{2 H_1 \gamma}{\kappa ^{2}} \left(t_0 - \phi\right)^{\gamma -1}\ ,\qquad
V(\phi) = \frac{1}{\kappa ^{2}} \left[ \left( 3H_0
+ 3H_1 \left(t_0 - \phi\right)^\gamma \right)^2
+ H_1 \gamma \left(t_0 - \phi\right)^{\gamma -1} \right]\ .
\ee
Thus, with such potential choice, the singularity (\ref{fs1}) could be
easily realized and the
energy density is, of course, finite.
It is easy to construct the models showing the finite future singularity
in other models, say, the scalar-Gauss-Bonnet theory,
non-minimal theories\cite{bamba}, etc.

We now consider what kind of scalar potential can generate the singularity in the
scalar field in the flat space-time background.
As a form of the singularity, we now assume
\be
\label{ss1}
\phi = \phi_0 + \phi_1 \left(t_0 - t\right)^\gamma\ .
\ee
Here $\phi_0$, $\phi_1$, and $\gamma$ are constants. One may take
$\gamma$ to be positive but not
integer. Then some derivative of the scalar field has singularity.

In the next section, we will rewrite the $F(R)$-gravity in the scalar-tensor form,
where  the metric is rescaled as given in (\ref{JGRG22}). Due to the scale
transformation,
the curvature singularity in the original frame of the $F(R)$-gravity does not appear
in the rescaled metric in the  the scalar tensor frame. The singularity
occurs
in the scalar field as follows from  (\ref{ss1}).
Hence, such scalar field  singularity (\ref{ss1}) is called finite-time
singularity even in flat Minkowski space.

In the flat space-time background, the equation of the scalar field is given by
\be
\label{ss2}
\ddot \phi + V'(\phi) = 0\ .
\ee
Here $V(\phi)$ is a potential of the scalar field.
By substituting (\ref{ss1}) into (\ref{ss2}), we find
\be
\label{ss3}
\phi_1 \gamma \left(\gamma - 1\right)\left(t_0 - t\right)^{\gamma -2} + V'(\phi) = 0\ .
\ee
Since (\ref{ss1}) can be rewritten as
\be
\label{ss4}
t_0 - t = \left(\frac{\phi - \phi_0}{\phi_1}\right)^{1/\gamma}\ ,
\ee
by substituting the expression (\ref{ss4}) into (\ref{ss3}), we find
\be
\label{ss5}
\phi_1 \gamma \left(\gamma - 1\right) \left(\frac{\phi - \phi_0}{\phi_1}\right)^{1 - 2/\gamma}
+ V'(\phi) = 0\ ,
\ee
which gives the form of the potential as
\be
\label{ss6}
V(\phi) = V_0
 - \frac{\phi_1 \gamma^2}{2} \left(\frac{\phi - \phi_0}{\phi_1}\right)^{2 - 2/\gamma}\ .
\ee
Hence, it is found that finite-time singularity  (\ref{ss1}) can be
realized by the potential (\ref{ss6}) even in flat space-time.
Thus, we demonstrated that for variety of dark energy models including
modified gravity the
finite-time singularity easily occurs even in the situation when effective
EoS parameter is bigger than $-1$ (the effective  quintessence).

\

\noindent
\section{The avoidance of finite-time singularity in modified $F(R)$-gravity}
\label{3}

Let us consider the action  (\ref{fr1}) again.
By introducing the auxiliary field $A$, we rewrite the action (\ref{fr1}) of the $F(R)$-gravity
in the following form:
\be
\label{JGRG21}
S=\frac{1}{2\kappa^2}\int d^4 x \sqrt{-g} \left\{F'(A)\left(R-A\right) + F(A)\right\}\ .
\ee
Here we neglect the contribution from the matter.
By the variation over $A$, one obtains $A=R$. Substituting $A=R$ into
the action (\ref{JGRG21}),
one can reproduce the action in (\ref{fr1}). Furthermore, we rescale the
metric in the following way (conformal transformation)\cite{NO}:
\be
\label{JGRG22}
g_{\mu\nu}\to \e^\sigma g_{\mu\nu}\ ,\quad \sigma = -\ln F'(A)\ .
\ee
Hence, the Einstein frame action is obtained:
\bea
\label{JGRG23}
S_E &=& \frac{1}{2\kappa^2}\int d^4 x \sqrt{-g} \left( R - \frac{3}{2}g^{\rho\sigma}
\partial_\rho \sigma \partial_\sigma \sigma - V(\sigma)\right) \ ,\nn
V(\sigma) &=& \e^\sigma g\left(\e^{-\sigma}\right)
 - \e^{2\sigma} f\left(g\left(\e^{-\sigma}\right)\right) = \frac{A}{F'(A)} - \frac{F(A)}{F'(A)^2}
\eea Here $g\left(\e^{-\sigma}\right)$ is given by solving the
equation $\sigma = -\ln\left( 1 + f'(A)\right)=\ln F'(A)$ as
$A=g\left(\e^{-\sigma}\right)$. In terms of $f(R)$ the potential
$V(\sigma)$ could be rewritten as \be \label{f(R)1}
V(\sigma)=\frac{A f'(A) - f(A)}{\left(1 + f'(A)\right)^2}\ . \ee
For the class of models with $f(R)$ behaving as in (\ref{frlv3}),
we find \be \label{fff1} V(\sigma) \sim -2\Lambda -
\frac{(n+1)\alpha}{R^n}\ , \ee for large scalar curvature. When
the curvature is large in the model (\ref{frlv3}), one gets \be
\label{fff2} \sigma \sim \frac{\alpha n}{R^{n+1}}\ . \ee By
combining (\ref{fff1}) and (\ref{fff2}), the potential $V(\sigma)$
is given in terms of $\sigma$ as: \be \label{fff3} V(\sigma) \sim
-2\Lambda - (n+1)\alpha\left(\frac{\sigma}{\alpha
n}\right)^{n/(n+1)}\ . \ee By comparing (\ref{fff1}) with
(\ref{ss6}) and identifying $\sigma$ with $(\phi -
\phi_0)/\phi_0$, we find \be \label{fff4} \frac{n}{n+1} = 2 -
\frac{2}{\gamma}\ , \ee or \be \label{fff5} \gamma =
\frac{2(n+1)}{n+2}\ . \ee Since $\gamma$ is fractional in general,
the scalar field $\sigma$ generates the singularity in
(\ref{ss1}), which now corresponds to the curvature singularity in
the Jordan frame. We should note that the curvature in the
Einstein frame is in general not singular.

Let us consider the realistic models which  unify the
early-time inflation
and late-time acceleration and which were introduced in refs.
\cite{Nojiri:2007as, Nojiri:2007cq}.

In order to construct such models,  the following conditions are used:
\begin{itemize}
\item Condition  that  inflation  occurs:
\be
\label{JGRG41}
\lim_{R\to\infty} f (R) = - \Lambda_i\ .
\ee
Here $\Lambda_i$ is an effective early-time cosmological constant.
\item The condition that there is flat space-time solution is given
as
\be
\label{JGRG44}
f(0)=0\
\ee
\item The condition that late-time acceleration occurs should be
\be
\label{JGRG45}
f(R_0)= - 2\tilde R_0\ ,\quad f'(R_0)\sim 0\ .
\ee
Here $R_0$ is the current curvature of the universe and we assume $R_0> \tilde R_0$.
Due to the condition (\ref{JGRG45}), $f(R)$ becomes almost constant in the present universe
and plays the role of the effective small cosmological constant:
$\Lambda_{\rm eff} \sim - f(R_0) = 2\tilde R_0$.
\end{itemize}
An example which satisfies the conditions  (\ref{JGRG41}),
(\ref{JGRG44}), and (\ref{JGRG45}) is given by the following action\cite{Nojiri:2007as}:
\be
\label{JGRG46}
f(R) = - \frac{\left(R-R_0\right)^{2n+1} + R_0^{2n+1}}{f_0
+ f_1 \left\{\left(R-R_0\right)^{2n+1} + R_0^{2n+1}\right\}}\ .
\ee
Here $n$ is a positive integer.
%
%

The conditions (\ref{JGRG41}) and (\ref{JGRG45}) require
\be
\label{JGRG47}
\frac{R_0^{2n+1}}{f_0 +f_1  R_0^{2n+1}}=2\tilde R_0\ ,\quad
\frac{1}{f_1} = \Lambda_i\ .
\ee
One can show that for the above class of models, there does not occur the
singularity or it
is difficult that the singularity occurs in visible future.
We now work in the scalar-tensor form or the Einstein frame  (\ref{JGRG23}).

In case of the model in \cite{HS}, in the limit of $A=R\to
+\infty$, $f(A)$ becomes a small constant $-\Lambda_{\rm eff}$
corresponding to the small effective cosmological constant in the
present universe. In the limit, we find $f'(A)\to 0$ and the
potential (\ref{f(R)1}) behaves as in (\ref{fff1}) and therefore
\be \label{f(R)2} V(\sigma) \to \Lambda_{\rm eff}\ . \ee Then the
value of $V$ is  very small and the curvature singularity could be
easily generated. In the models exhibiting this kind of
singularity due to fact that singularity is at finite distance
with current energy scale, there was conjectured
\cite{KobayashiMaeda} that neutron stars cannot be formed. (This
is really conjecture because the derivation of stars formation
follows the same approximation as in the usual Einstein gravity,
actually neglecting the higher derivatives terms typical in
$F(R)$-gravity. This derivation should be reconsidered within the
real $F(R)$-gravity with account of higher derivative terms as
well as non-linearity what is technically not easy task. For
instance, the non-linear structure of modified gravity leads to
oscillations\cite{baty} already in the very simple approximation.)
The conjecture is that if  the density of the matter becomes
finite but large and reaches a critical value, the curvature
singularity occurs and the star could collapse. Therefore the star
with density larger than the critical value could not be formed.
As is mentioned above these simple considerations may be not valid
with account of higher derivative terms where star formation
process should be reconsidered from the very beginning.

On the other hand, for the class of models satisfying the conditions (\ref{JGRG41}-\ref{JGRG45}) like
the model (\ref{JGRG47}),
the singularity does not occur easily. For this class of the models,
$f'(R)$ vanishes at $R=R_0$ and
in the limit of $R\to \infty$.
When $R=R_0$, the condition (\ref{JGRG41}) gives
\be
\label{f(R)3}
V(\sigma)\to 2\tilde R_0\ ,
\ee
which could be quite small. Since the value of $R$ is finite,
(\ref{f(R)3})
does not correspond to any singularity. On the other hand, in the limit of $R\to +\infty$,
by using the condition (\ref{JGRG45}), we find
\be
\label{f(R)4}
V(\sigma)\to \Lambda_i\ .
\ee
Since $\Lambda_i$ corresponds to the effective cosmological constant
during the inflation, the energy
scale is not small, typically it is the Grand Unification scale.
Therefore the value of $V(\sigma)$ could be very large. Then the singularity
could be generated only at the energy density larger then the energy
density corresponding to the inflation
of the early universe but it does not occur around the energy-density
which is typical  for neutron star.

Even for the class of dark energy models where singularity occurs at
smaller energies, there is scenario to avoid the singularity proposed in
ref.\cite{Abdalla}. Indeed, let us consider  the model
where $f(R)$ behaves as \be
\label{ans}
f(R)\sim f_0 R^\alpha\ ,
\ee
with constants $f_0$ and $\alpha>1$. If the ideal fluid,
which could be the matter
with the constant EoS parameter $w$: $p=w\rho$, couples with the gravity,
when the $f(R)$-term dominates compared with the Einstein-Hilbert
term, an exact solution is \cite{Abdalla}
\bea
\label{ans2}
&& a=a_0 t^{h_0} \ ,\quad h_0\equiv \frac{2\alpha}{3(1+w)} \ ,\nn
&& a_0\equiv \left[-\frac{6f_0h_0}{\rho_0}\left(-6h_0 + 12 h_0^2\right)^{\alpha-1}
\left\{\left(1-2\alpha\right)\left(1-\alpha\right) - (2-\alpha)h_0\right\}\right]^{-\frac{1}{3(1+w)}}\ .
\eea
When $\alpha=1$, the result $h_0 = \frac{2}{3(1+w)}$ in the Einstein gravity is reproduced.
The effective EoS parameter $w_{\rm eff}$ may be defined by
\be
\label{ans3}
h_0=\frac{2}{3\left(1+w_{\rm eff}\right)}\ .
\ee
By using (\ref{ans2}), one finds
\be
\label{M9}
w_{\rm eff}=-1 + \frac{1+w}{\alpha}\ .
\ee
Hence, if $w$ is greater than $-1$ (effective quintessence or even usual ideal fluid with positive $w$),
when $\alpha$ is negative, we obtain the effective phantom phase where $w_{\rm eff}$ is less than $-1$.
This is different from the case of pure modified gravity.
On the other hand, when $\alpha>w+1$ (it can be even positive),  $w_{\rm eff}$ could
be negative for negative $w$.
Hence, it follows that modified gravity minimally coupled with usual (or quintessence) matter
may reproduce quintessence (or phantom) evolution phase for dark energy universe
in an easier way than without such coupling.

If we choose $\alpha$ to be negative in (\ref{ans}), when the curvature is small,
the $f(R)$ term becomes dominant compared with the Einstein-Hilbert term.
Then from (\ref{M9}), we have an effective phantom even if $w>-1$. Usually
the phantom generates the Big Rip singularity.
However, near the Big Rip singularity, the curvature becomes large and
the Einstein-Hilbert term becomes dominant. In this case, we have $w_{\rm eff}\sim w >-1$,
which prevents the Big Rip singularity.

One may add extra term to $f(R)$ (\ref{ans}) as \cite{Abdalla, bamba}.

%
%

\be
\label{ans4}
f(R) =  f_0 R^\alpha + f_1 R^\beta \ ,
\ee
Here we choose $f_1>0$ and $\beta>1$. Then for the large curvature, the second term could
dominate. Then for the large curvature, the potential (\ref{f(R)1}) behaves as
\be
\label{ans5}
V = \frac{\beta-2}{f_1 R^{\beta -2}}
\ee
%
%
Then if $1<\beta<2$, the potential is positive and diverges near the curvature singularity
$R\to \infty$, which could prevent the curvature singularity even if the
singularity is
 the Big Rip type (Type I) or Type II\cite{bamba} or other softer
singularity\cite{bamba}.

Thus, we demonstrated that for large class of viable $F(R)$-gravities the
finite-time singularity occurs in so distant future that it cannot
influence the current universe processes. From another side, there exists
the trick introduced in refs.\cite{Abdalla, bamba} how by adding extra
term to modified gravity to remove the singularity.

\noindent
\subsection{Singularity avoidance in  integral modified gravity models}

As it is clear from (\ref{JGRG21}), for $F'(R)=1+f'(R)>0$, the
square of the effective gravitational coupling is positive.
However, due to $\kappa_{\rm eff}^2 \equiv \kappa^2 / F'(A)$, the
theory could enter an anti-gravity regime. In order to avoid this
anti-gravity problem from the very beginning, we may consider a
model given by \be \label{Uf16} f(R)=-f_0 \int_0^R dR
\e^{-\frac{\alpha R_1^{2n}}{\left(R - R_1\right)^{2n}}
 - \frac{R}{\beta\Lambda_i}}\ .
\ee
Here $\alpha$, $\beta$, $f_0$, and $R_1$ are constants and we assume
$0<R_1\ll \Lambda_i$.
In this model, the correction to the Newton law is very small.
Then by construction, as long as $0<f_0<1$, we find $f'(R)>-1$ or $F'(R)>0$
and therefore there is no anti-gravity problem.
Since
\be
\label{Uf17}
f(R_1) \sim -f_0 \int_0^{R_1} dR \e^{-\frac{\alpha R_1^{2n}}{\left(R - R_1\right)^{2n}}}
= -f_0 A_n(\alpha) R_1\ , \quad
A_n(\alpha) \equiv \int_0^1 dx \e^{-\frac{\alpha}{x^{2n}}}\ ,
\ee
and $- f(R_1)$ could be identified with the effective cosmological
constant $2\tilde R_0$, we find
\be
\label{Uf18}
f_0 A_n(\alpha) R_1 = \tilde R_0\ .
\ee
Note that $A_n(0)=1$, $A_n(+\infty)=0$, and $A'(x)<0$.
On the other hand, since
\be
\label{Uf19}
f(+\infty) \sim \int_0^{\infty}dR \e^{-\frac{R}{\beta \Lambda_i}} = - f_0 \beta \Lambda_i\ ,
\ee
and $-f(+\infty)$ could be identified with the effective cosmological
constant at the inflationary epoch, $\Lambda_i$, we find
\be
\label{Uf19B}
f_0\beta=1\ .
\ee
Then the conditions (\ref{JGRG41}) and (\ref{JGRG45}) are satisfied. The condition (\ref{JGRG41})
is, of course, satisfied by construction (\ref{Uf16}).
As discussed around Eq.(\ref{f(R)4}), as long as $\Lambda_i$ could be large enough,
the potential becomes large when the scalar curvature $R$ is large and there could not occur or could
be difficult to realize the singularity.
We also note that if we add the second term $f_1 R^\beta$ with $1<\beta<2$ in (\ref{ans4}),
the singularity is completely removed since the potential diverges in the limit of $R\to \infty$.
Thus, for large class of viable modified gravity models the singularity
maybe easily removed by adding of extra term which is actually relevant at
the early universe. Moreover,the typical  energy scales of neutron star
and singularity
formation process (above the inflation scale) are at large distances and
they are not relevant for each other.

\section{The singularity problem in $f(R)$-viable-models}
\label{4}
In this section, we are going to discuss the  curvature
singularity problem that affects several infrared-modified $f(R)$
model\cite{HS}. As we have  said
above, considering $f(R)\neq R$-gravity means that a new scalar
degree of freedom has to be taken into account. Conformal
transformations of the metric can be used to make it explicit in
the action \cite{Whitt:1984pd, Maeda:1988ab}.

For our goals, we consider  a class of $f(R)$-models which do not
contain cosmological constant and  are explicitly designed to
satisfy cosmological and Solar-System constraints.  In practice,
we choose a class of functional forms of $f(R)$ capable of
matching, in principle, observational data \cite{mimic}. First of
all, any viable cosmological model  have to reproduce the CMBR
constraints in the high-redshift regime.  Secondly, it should give
rise to an accelerated expansion, at low redshift, according to
the $\Lambda$CDM model. Thirdly, these models should give rise to
a large mass for the scalaron \cite{Star80} in the high-density
region where local gravity experiments are carried out. In such a
regime, the perturbation in $R$ can be larger than the background
value $R_0$, which means that the linear approximation to derive
the Newtonian effective gravitational constant in a spherically
symmetric spacetime ceases to be valid \cite{Navarro,Erick}. In
the non-linear regime with a heavy scalaron mass, however, it is
known that a spherically symmetric body has a thin-shell
\cite{HS,Navarro,Faul,TUT} through so called the {\it chameleon
mechanism} \cite{chame1,chame2} (see also
Ref.~\cite{Mota,captsu}). When a thin-shell is formed, an
effective coupling that mediates the fifth force gets smaller.
This allows the possibility that the $f(R)$ models which have a
large scalaron mass in the high-density region can be compatible
with local gravity experiments. Then there should be sufficient
degrees of freedom in the parameterization to encompass low
redshift phenomena (e.g. the large scale structure) according to
the observations \cite{huspe}. Finally, small deviations from GR
should be consistent with Solar System tests. All these
requirements suggest that we have to assume the  limits
\begin{equation}
\lim_{R\rightarrow\infty}f(R)={\rm constant},
\end{equation}
\begin{equation}
\lim_{R\rightarrow0}f(R)=0,
\end{equation}
which are satisfied by a general class of broken power law models,
as those proposed in \cite{HS}

\begin{equation}
f_{HS}(R)=-\lambda
R_{c}\frac{\left(\frac{R}{R_{c}}\right)^{2n}}{\left(\frac{R}{R_{c}}\right)^{2n}+1}\,,
\label{eq:HS}\end{equation}

Since $f(R=0)=0$, the cosmological constant has to disappear in a
flat spacetime. The parameters  $\{n$, $\lambda$, $R_{c}\}$
are constants which should be determined by experimental bounds.

Other interesting models with similar features  have been studied
in \cite{Nojiri:2007as,Appleby,Tsuji,Nojiri:2007cq}. In all these
models, a de-Sitter stability point, responsible for the late-time
acceleration, exists for $R=R_1~(>0)$, where $R_1$ is derived by
solving the equation $R_1 f_{,R}(R_1)=2f(R_1)$ \cite{ottewill}.

In the region $R \gg R_c$,  model (\ref{eq:HS})
 behaves as
\begin{eqnarray}
\label{eq:fR} f_{hybrid1}(R) \simeq -\lambda R_c \left[1- \left( R_c/R
\right)^{2s} \right]\,,
\end{eqnarray}
where $s$ is a positive constant. The model approaches
$\Lambda$CDM  in the limit $R/R_c \to \infty$.

Finally, let  also consider the class of models
\cite{Li,AT07,Faul}

\begin{equation}
f_{hybrid2}(R)=-\lambda
R_{c}\left(\frac{R}{R_{c}}\right)^{q}\,.\label{eq:TS}
\end{equation}
Also in this case $\lambda$, $q$ and $R_c$ are positive constants
(note that $n$, $s$ and $q$ have to converge toward the same
values to match the observations). We do not consider the models
whit negative $q$, because they suffer for instability problems
associated with negative $f_{,RR}$ \cite{bean-f(R)}.

\subsection{The role of conformal transformations}
\begin{figure}
\begin{tabular}{|c|clclcl}
\hline
\includegraphics[scale=0.8]{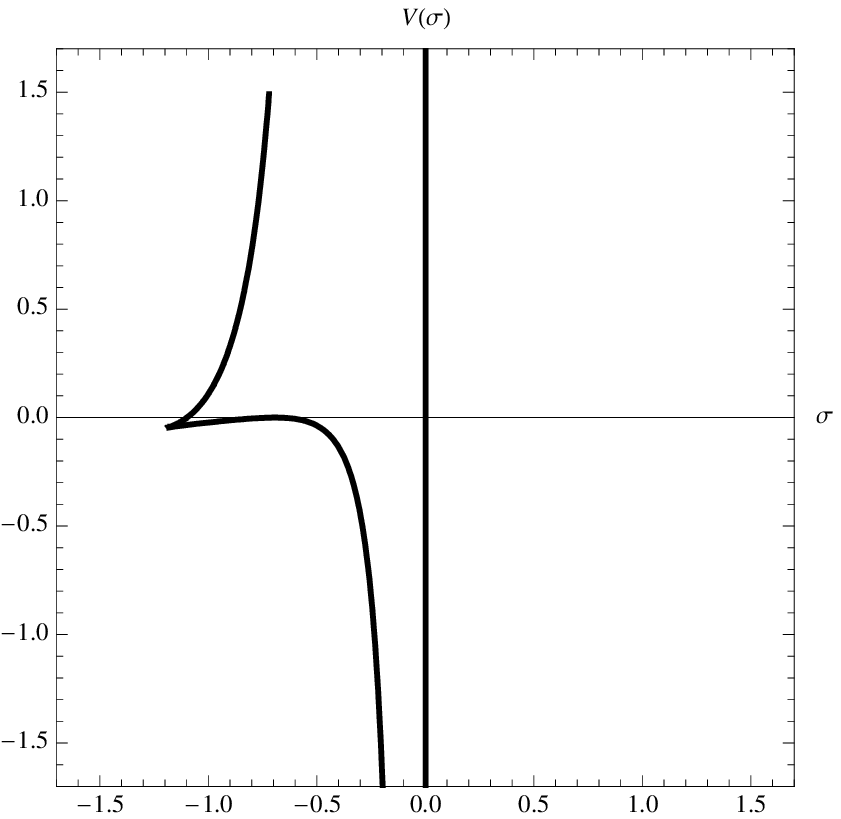}&
\includegraphics[scale=0.7]{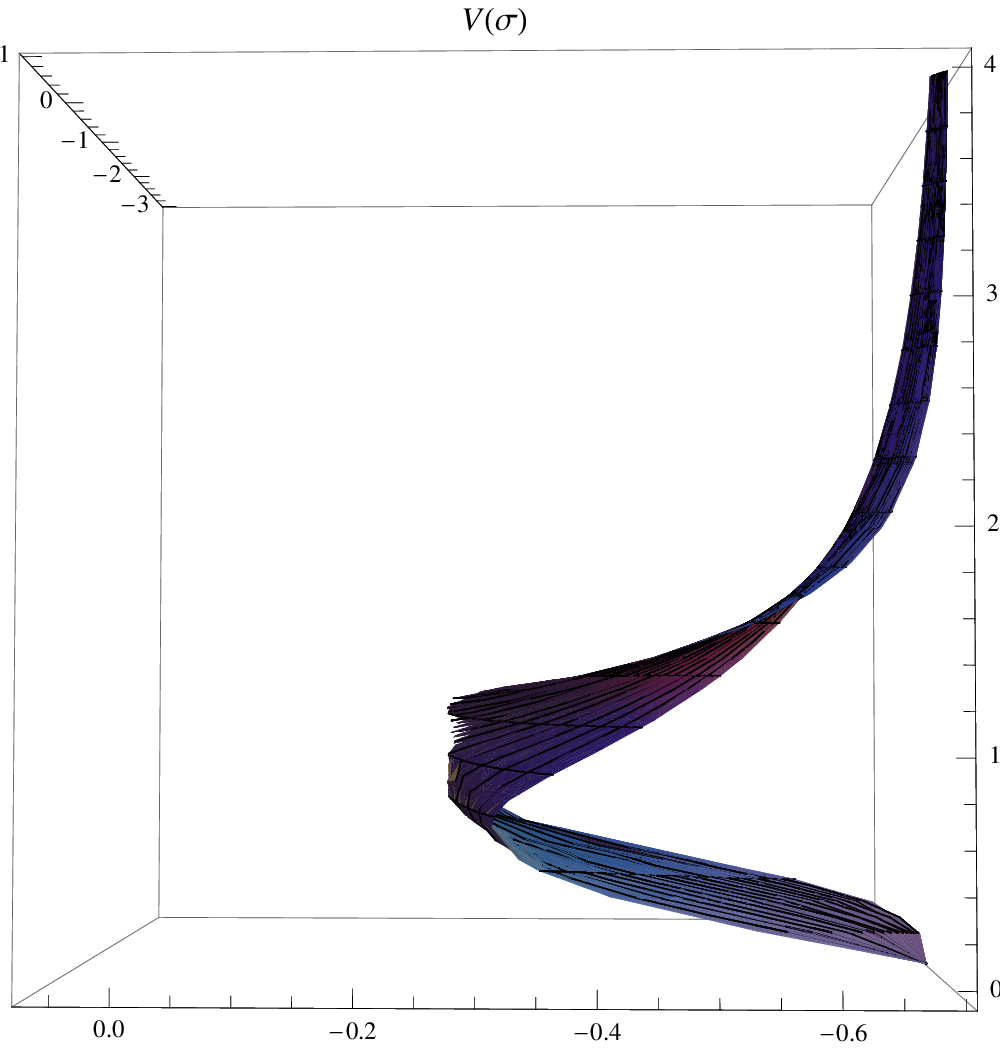}&\\
\hline
\end{tabular}
\caption {Qualitative evolution of the scalar degree of freedom
for the model  (\ref{eq:HS}) in two  dimensions
$\{V(\sigma),\sigma\}$ and in three dimension
$\{V(\sigma),\sigma,x\}$ with $\lambda=2$ and $n=1$.}
\label{fig:1}
\end{figure}

Conformal transformations can play a key role to classify
singularities in modified gravity (apart from general classification
presented in second section). As we have shown in above section, at scale
$R_c$ and beyond, the expansion rate of the Universe is set primarily by
the matter density, just as in standard cosmology, with small
corrections. Once the local curvature drops below $R_c$, according
to the {\it chameleon mechanism}, the expansion rate  feels the
effect of modified gravity. The spacetime curvature, on the other
hand, is controlled by the further scalar degree of freedom
$\sigma$ which gravity acquires. It obeys the usual scalar field
equation with potential $V(\sigma)$, the shape of which is
directly determined by function $f(R)$, and a driving term from
the trace of matter stress-energy tensor. But a problem comes out
at this point: it turns out that precisely those functions $f(R)$
that lead to Einstein-like gravity action in the large curvature
regime, yield a potential $V$ with an {\it unprotected} curvature
singularity. (Note that just just the same occurs for number of realistic
quintessence dark enrgies.) Let us consider, for example, the  model
(\ref{eq:HS}) which has been constructed to avoid  linear
instabilities. The conformal transformation gives $\sigma=-\ln
(1+f'(R))=-\ln F'(R)$ with the potential defined  in
Eq.(\ref{f(R)1}).  For the model (\ref{eq:HS}), the scalar field
$\sigma$ is given by
\begin{equation}\label{eq:HSphi}
 \sigma(x) =-\ln \left(2-\frac{2 n x^{2 n-1} \lambda }{\left(x^{2
   n}+1\right)^2}\right)
 \end{equation}
where the cross-over curvature scale $R_c$ can be reabsorbed  into a
rescaling of coordinate  (which can be assumed dimensionless and
can be  measured in length units corresponding to $R_c$). Suitable
coordinates are $x=\frac{R}{R_c}>0$ and $R_c\sim \rho_g \sim
10^{-24}$\,g/cm$^3$ for the  Galactic density in the Solar
vicinity and $R_c\sim \rho_g \sim 10^{-29}$\,g/cm$^3$ for the
present cosmological density.
For the large curvature limit $R \rightarrow \pm \infty$, we have
that $\sigma=-0.70$ while, for $R \rightarrow0$, it  corresponds
to $\sigma=-0.70$, which gives us a hint that the potential is
going to be a multi-valued function. For $R\rightarrow 1$, we have
$\sigma=0$. We have
$V(\sigma)\rightarrow\infty$  for $R\rightarrow1$, then it is
singular for this  curvature value, while for $R\rightarrow
\infty$, we have $V(\sigma)=2$ and flat spacetime in the limit
$R\rightarrow 0$. The analytic expression is

%
%
\begin{equation}
V(\sigma(x))=\frac{x^2 \left(x^{2 n}-2 n+1\right) \left(x^n+x^{3
n}\right)^2 \lambda
   }{\left(2 (x-n \lambda ) x^{2 n}+x^{4 n+1}+x\right)^2}\,.
   \label{eq:HSpotential}
   \end{equation}
%
%
For our goals, it is important to study the trend of the potential
and of the field $\sigma$ through a parametric plot of two
functions with respect to $x=\frac{R}{R_c}$. In this way,
singularities can be easily identified (see Fig. \ref{fig:1}).

For the hybrid-1 model, $\sigma$ assumes the form

\begin{equation}\label{eq:H1sigma}
\sigma=-\ln \left(2-2 s \left(\frac{1}{x}\right)^{2 s+1} \lambda
\right)\,.
\end{equation}

\begin{figure}
\centering
\includegraphics[scale=0.5]{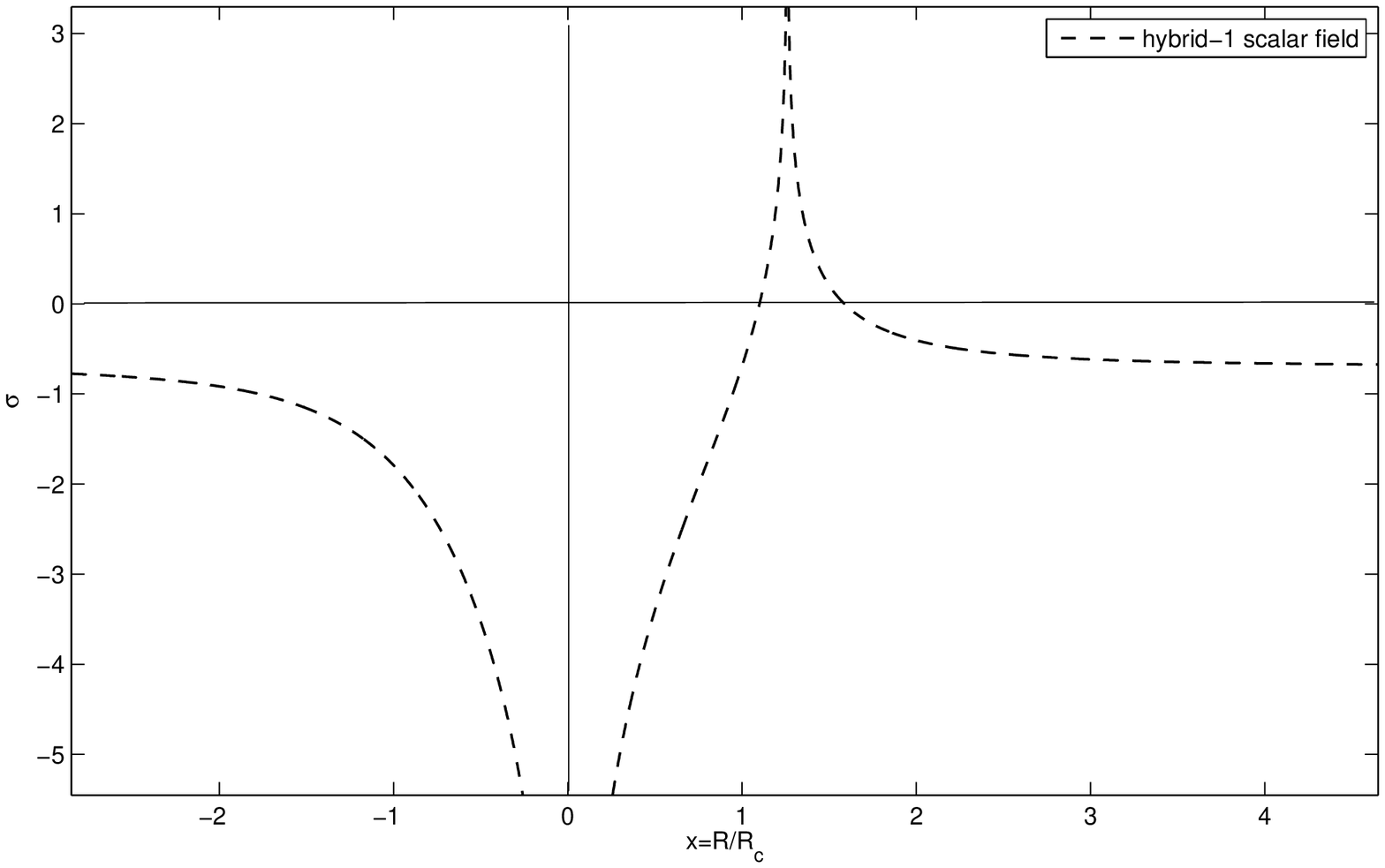}
\caption{Plot of  Eq.(\ref{eq:H1sigma}) for $\lambda=2$ and
$s=1$.} \label{fig:HYB1scalarfield}
\end{figure}

>From Fig.\ref{fig:HYB1scalarfield}, for $\lambda=2$ and $s=1$, we
can see that we have $\sigma=0$ in the limit $R \rightarrow -
\infty$ and $\sigma \rightarrow\infty$ for $R= 0$.

\begin{figure}
\centering
\includegraphics[scale=0.5]{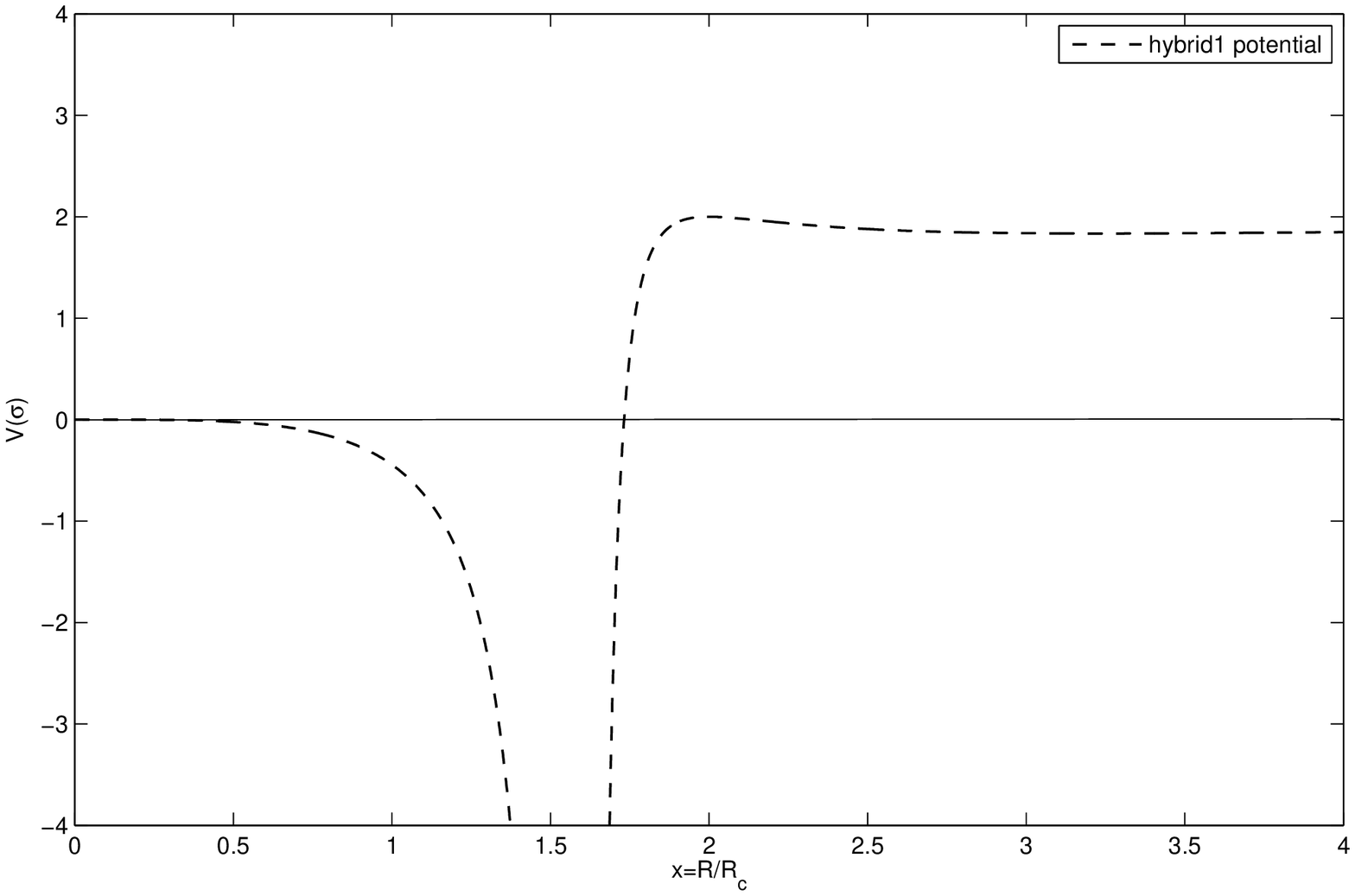}
\caption{Plot of  Eq.(\ref{eq:H1potential}) for $\lambda=2$ and
$s=1$.} \label{fig:H1potential}
\end{figure}

The effective potential  is
\begin{equation}
\label{eq:H1potential}
V(\sigma)=\frac{x^{2 s+2} \left(x^{2 s}-2 s-1\right) \lambda }{\left(x^{2 s+1}-2
   s \lambda \right)^2}\,,
   \end{equation}
and, from a rapid inspection of  Fig. \ref{fig:H1potential} with
$\lambda=2$ and $s=1$, we have $V=2$ for $R\rightarrow\pm\infty$.

Finally, let us analyze  the model (\ref{eq:TS}). The scalar field
assumes the form
\begin{equation}\label{eq:phyhybridtwo}
\sigma=-\ln \left(q \lambda
\left(\frac{1}{x}\right)^{q+1}+2\right)\end{equation} and then
$\sigma\rightarrow-\infty$ for $R \rightarrow 0$.
\begin{figure}
\centering
\includegraphics[scale=0.5]{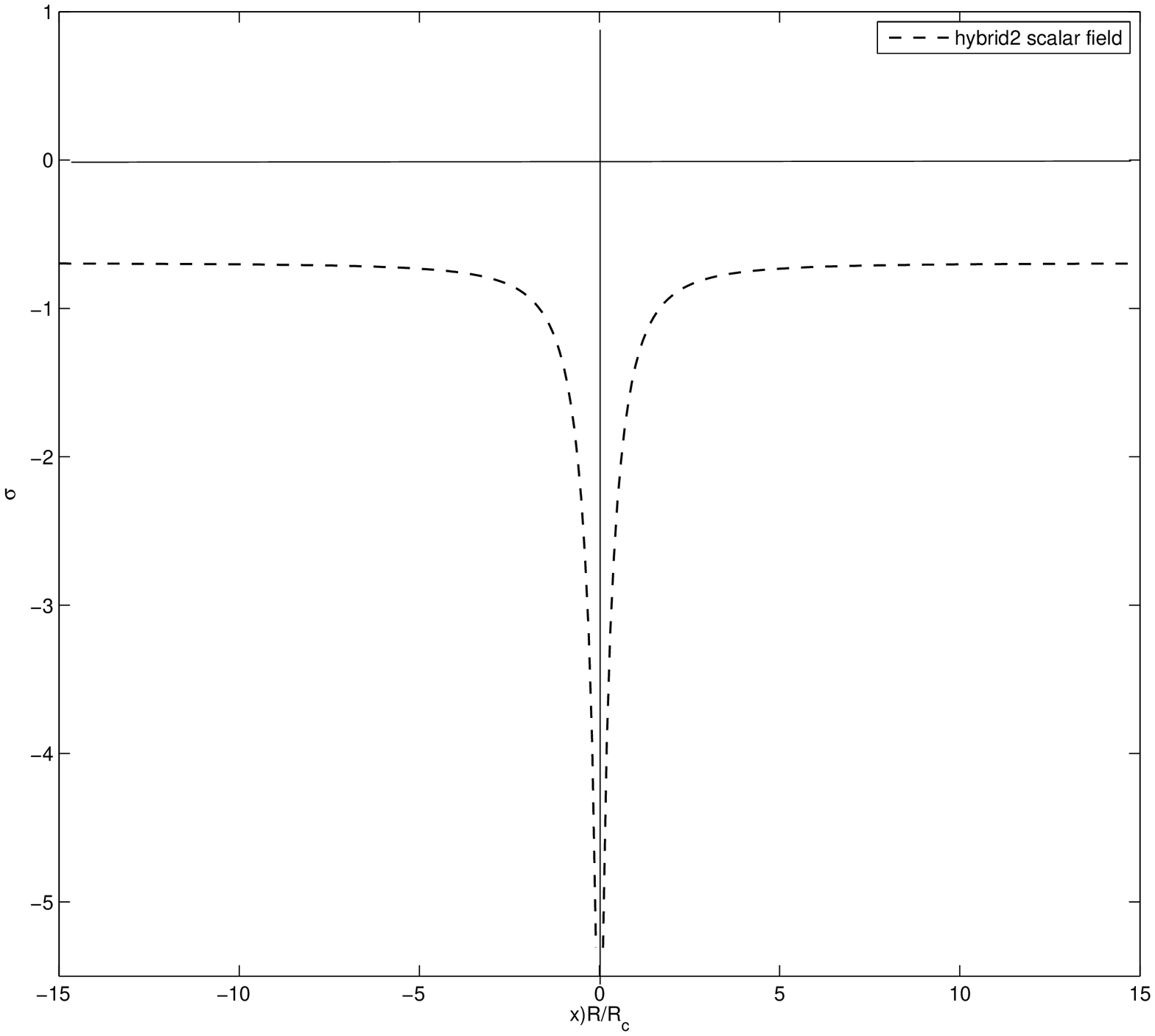}
\caption{Plot of  Eq.(\ref{eq:phyhybridtwo}) for $\lambda=2$ and
$q=1$.} \label{fig:HYB2scalarfieldl}
\end{figure}

The potential is
\begin{equation}\label{eq:H2potential}
V(\sigma)=\frac{(q+1) x^{q+2} \lambda }{\left(x^{q+1}+q \lambda \right)^2}
\end{equation}
which gives $V=0$ for $R \rightarrow\pm \infty$ and for $R=0$.
\begin{figure}
\centering
\includegraphics[scale=0.5]{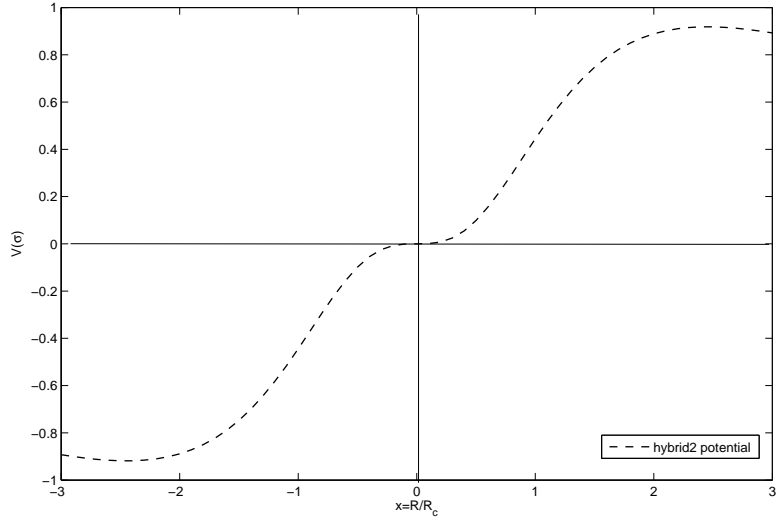}
\caption{Plot of  Eq.(\ref{eq:H2potential}) for $\lambda=2$ and
$q=1$, it is worth of noting the double sigmoid trend.}
\label{fig:H2potential}
\end{figure}
\begin{figure}
\centering
\includegraphics[scale=0.5]{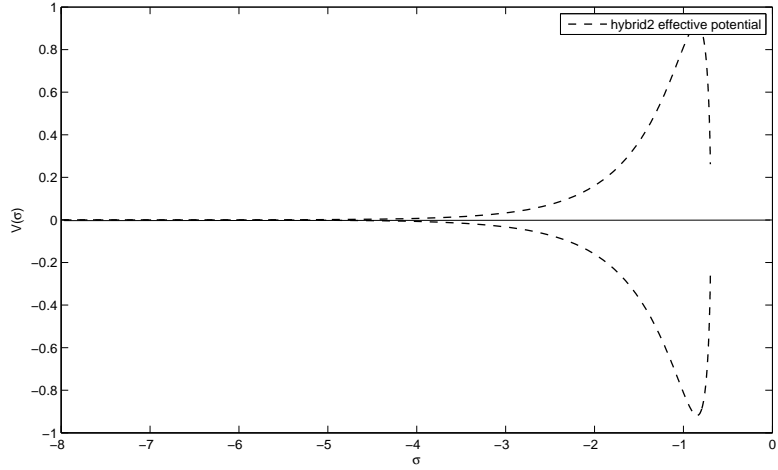}
\caption{Parametric plot of  Eqs.(\ref{eq:H2potential}) and
(\ref{eq:phyhybridtwo}) for $\lambda=2$ and $q=1$.}
\label{fig:HB2parametric}
\end{figure}
As concluding remark, we can say that the conformal
transformations allow to recast the singularity problem in terms
of the scalar field and its potential. This fact suggests an easier
interpretation of the singularities, at least, in some specific cases.

\subsection{Singularities vs Chameleon mechanism}

In order to compare these results,  we can take into account
another approach where the  matter coupling is considered as also
been studied  already from Sami et al. \cite{Dev}.  This allows to
compare singularities with density scale and could be particularly
useful to construct  models where finite singularities  are
avoided at infrared scales.

Starting again with  action (\ref{fr1})  which leads to the
equations of motion (\ref{JGRG13}), the evolution of the scalar
degree of freedom is given by the trace:

\be 2F(R) - R F'(R) - 3 \Box F'(R) = - \frac{\kappa^2}{2} T\ . \ee
Another convenient way to define the scalar function $\sigma$ is
\begin{equation}
 \sigma \equiv F'(R) - 1,
\end{equation}
 which is expressed through Ricci scalar once $F(R)$  is specified.
We can write the trace equation (equation $(\ref{frlv1})$)
 in the term of $V$ and $T$ as
\begin{equation}
 \Box \sigma = \frac{dV}{d\sigma} +  \frac{\kappa^2}{6} T\ .
\end{equation}
\vskip 0.5cm
\noindent The potential can be evaluated using the following relation
\begin{equation}\label{Effpot}
 \frac{dV}{dR} = \frac{dV}{d\sigma}\frac{d\sigma}{dR}= \frac{1}{3}
\left ( 2 F(R) - F'(R) R \right ) F''(R).
\end{equation}
then $V(\sigma)$ is  given by the pair of functions
$\{\sigma(R),V(R)\}$.
Let us consider now the model
(\ref{eq:HS}) as also is showed in \cite{Dev}. The scalar field $\sigma$ is given by
\begin{equation}\label{eq:phi}
 \sigma(R) = -\frac{2 n x^{2 n-1} \lambda }{\left(x^{2 n}+1\right)^2}.
\end{equation}
We can compute $V(R)$ for a given value of $n$. In the case of $n
= 1$ and $\lambda=1$, we have
\begin{equation}
V=\frac{1}{24} \left(\frac{-3 x^7-24 x^6+21 x^5-56 x^4+11 x^3-40 x^2+3
   x-8}{\left(x^2+1\right)^4}-3 \tan ^{-1}(x)\right)\label{potentialhs}
   \end{equation}

\noindent where $x=R/R_c$. In the FRW background, the trace
equation can be rewritten in the convenient form
\begin{equation}\label{eq:scalarfield}
 \ddot{\sigma} + 3H\dot{\sigma} + \frac{dV}{d\sigma} =  -\frac{\kappa^2}{6} \rho.
\end{equation}

\begin{figure}
\begin{tabular}{|c|clc}
\hline
\includegraphics[scale=0.8]{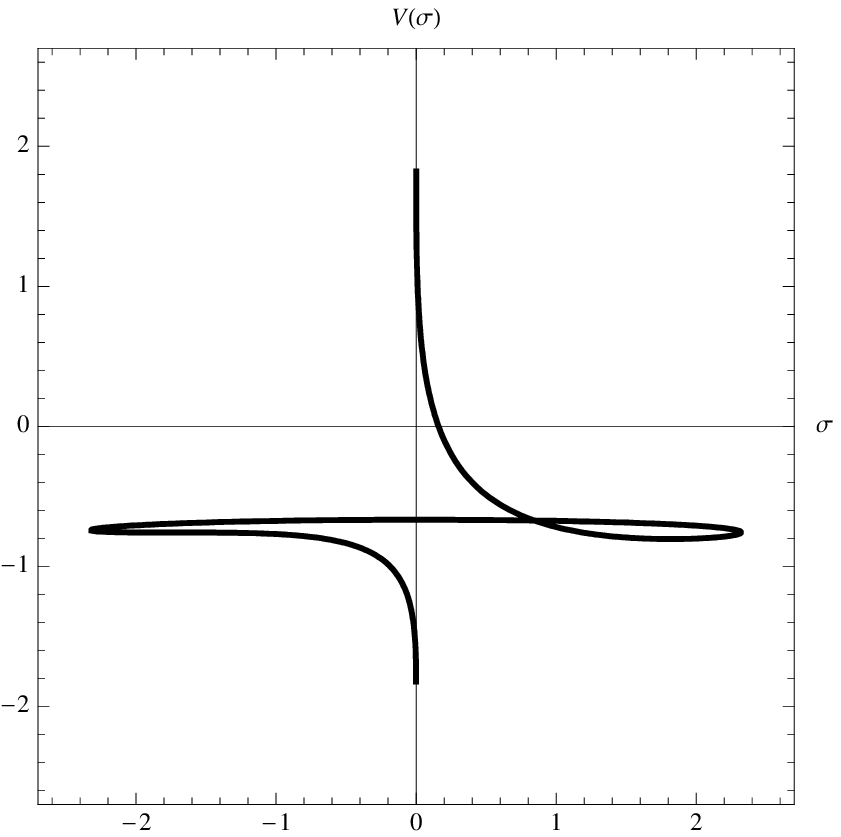}&
\includegraphics[scale=0.7]{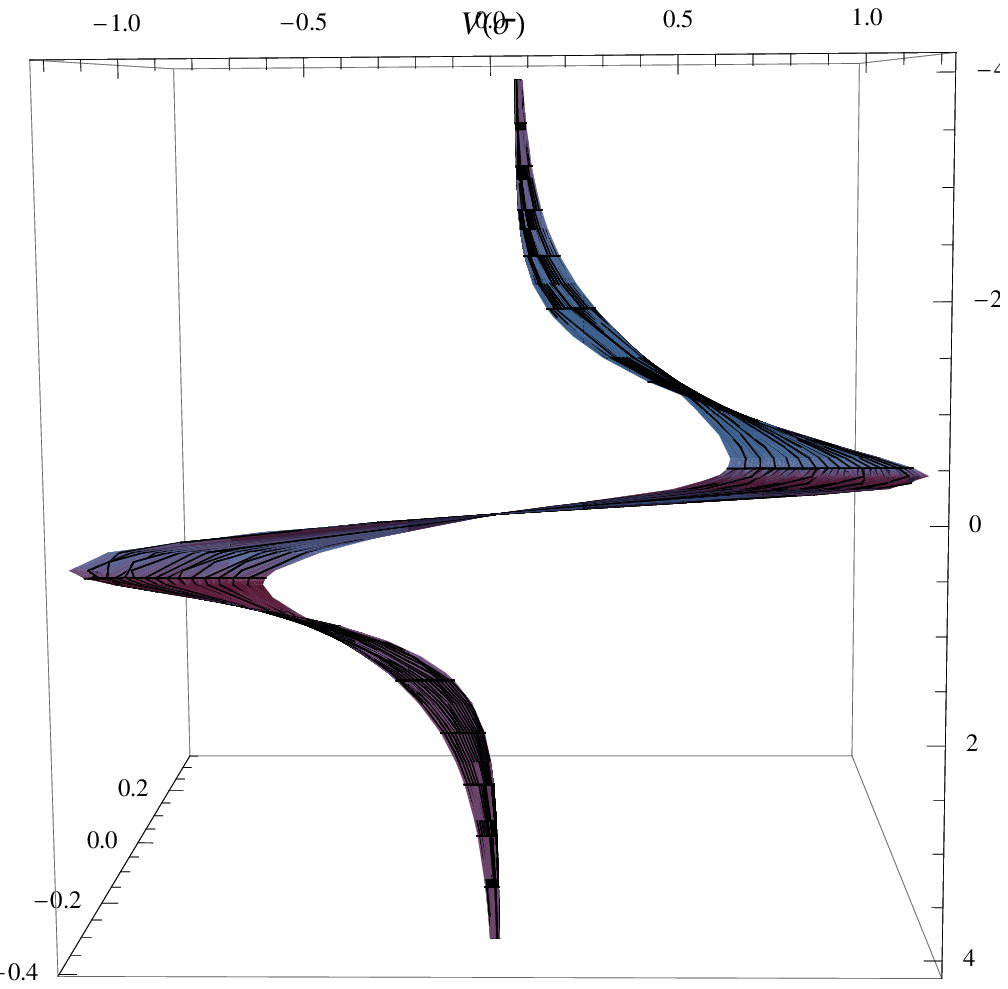}\tabularnewline
\hline
\end{tabular}
\caption {Qualitative parametric plot of the evolution of the
scalaron potential $V$ vs $\sigma$ and $x$ for $n =1$ and $\lambda
= 1$ in two and three dimensions for the model
(\ref{eq:HS}).} \label{fig:2}
\end{figure}
\begin{figure}
\begin{tabular}{|c|clc}
\hline
\includegraphics[scale=0.8]{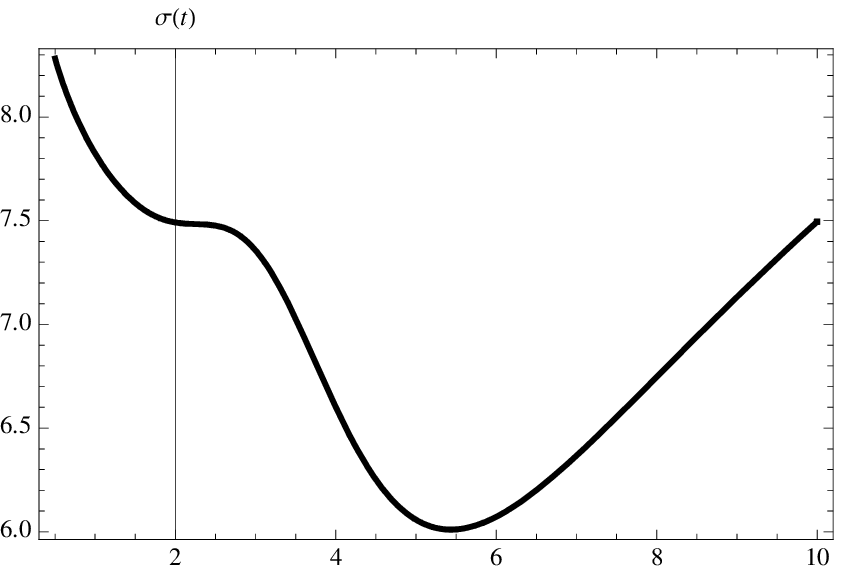}&
\includegraphics[scale=0.8]{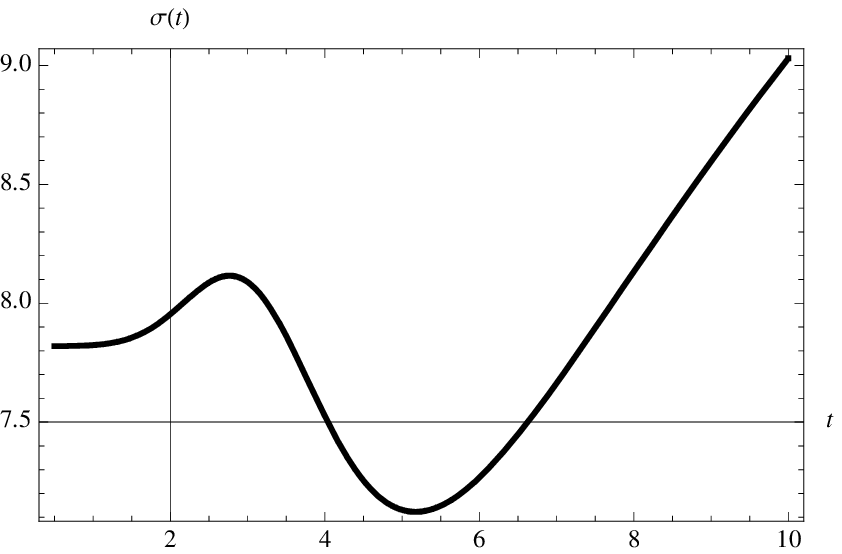}\tabularnewline
\hline
\end{tabular}
\caption {Cosmological behavior of the scalar field $\sigma$ as
given in Eq. (\ref{eq:scalarfield}) plotted vs the cosmological
time in presence of matter (left) and without matter (right). It
is possible to note that the matter contribution changes the local
behavior towards $t=0$.} \label{fig:matter}
\end{figure}

The effective scalar potential is plotted in Fig.~\ref{fig:2} for
$\lambda=1$, and it is  multivalued indeed. It has a minimum
depending on the values of $n$ and $\lambda$. For generic values
of the parameters, the minimum of the potential is close to
$\sigma =0$, corresponding to infinitely large curvature
$R=\pm\infty$. Thus, while the field is evolving towards the
minimum, it evolves oscillating towards a singular point. We have
a stable de Sitter minimum  and an unstable de Sitter maximum. The
point $R \rightarrow0$ corresponds to a flat spacetime, which,
although a solution for the model, is unstable. Cusps that occur
at $R=\pm 1/\sqrt{3}$ are critical points with $f''=0$. However,
depending upon the values of the parameters, we can choose a
finite range of initial conditions for which scalar field $\sigma$
evolves to the minimum of the potential without hitting the
singularity.   Note again, that as it was demonstrated in second
section, it is just manifestation of II Type finite-time
singularity. In Fig. \ref{fig:matter}, the cosmological behavior
of $\sigma$ is plotted. It is clear the contribution of matter in
changing the local behavior.

The time-time component of the equation of motion $(\ref{JGRG13})$
gives the Hubble equation
\begin{equation}\label{eq:Friedmann}
 H^2 + \frac{d (\ln F'(R))}{dt} H + \frac{1}{6}\frac{F(R) - F(R)R}{f'(R)}
= - \frac{\kappa^2}{6}  \rho.
\end{equation}
The Einstein gravity is recovered in the limit $f^{\prime }=1$ .
The  picture of dynamics which appears here is the following:
above infrared  scale $(R_{c})$, the expansion rate is set by the
matter density and once the local curvature falls below $R_{c}$
the expansion rate gets the effects of modified gravity.

For pressureless dust, the effective potential presents an extremum
at
\begin{equation}  \label{extm}
2 F(R) - R F'(R)=  - \frac{\kappa^2}{2} \rho\,.
\end{equation}
For  viable late time cosmologies, the field   evolves near the
minimum of the effective potential. The finite-time singularity,
which occurs for the class of models under consideration, severely
constrains the field dynamics.

\subsection{Adding higher curvature corrections to cure   singularities}

It is well-known that for large curvature regimes, the quantum
effects become important leading to higher curvature corrections.
Then the program of this paper can be enhanced by considering if
higher curvature corrections (as is already stressed in third section)
added to the original models can
solve the singularity problem.  In general, higher curvature
corrections change the structure of the effective potential around
the singularity \cite{Dev}. Keeping this in mind, let us consider
the modification of the model (\ref{eq:HS}).  Although we
focus on a specific model,  similar results hold for the models of
the classes considered here. In cosmology, higher curvature
corrections  appear  as Lagrangian contributions like ${\cal L
}=\alpha_2R^2+\alpha_3R^3+\cdots$, and so the most natural choice
for the leading order term is $\alpha R^2/R_{c}^2$.\footnote{
Higher order corrections naturally include terms like
$R_{\mu\nu}R^{\mu\nu}$, but in this paper we focus on the
$f(R)$-type modified gravity and hence simply assume that the
corrections are also given by a function of the Ricci scalar.} It
is well-known that the $R^2$ - term may be responsible for
inflation in the {\em early} Universe if $R_c$ is set to be at
inflationary scale \cite{Dev}. In the case which we are
considering, we have
\begin{equation}
F(R) = -\frac{\lambda  (\frac{R}{R_c})^{2 n}}{(\frac{R}{R_c})^{2 n}+1}+\alpha  \frac{R^2}{R_c^2}+\frac{R}{R_c},
\end{equation}
then the field $\sigma$ becomes
\begin{equation}\label{eq:phi2}
\sigma(R) = 2 \frac{R}{R_c} \alpha -\frac{2 n \lambda }{(\frac{R}{R_c})^{2 n+1}+\frac{R}{R_c}}+\frac{2 n \lambda }{\frac{R}{R_c}
   \left((\frac{R}{R_c})^{2 n}+1\right)^2}
\end{equation}

 When $|R|$ is large in modulus, the first term which comes from $\alpha R^2$
dominates. In this case, the curvature singularity $R = \pm \infty$
 corresponds to $\sigma = \pm \infty$. Hence, by this modification, the
minimum of the effective potential is separated from the curvature
singularity by the infinite distance in the $\{\sigma,
V(\sigma)\}$ plane.

\begin{figure}
\begin{tabular}{|c|clc}
\hline
\includegraphics[scale=0.8]{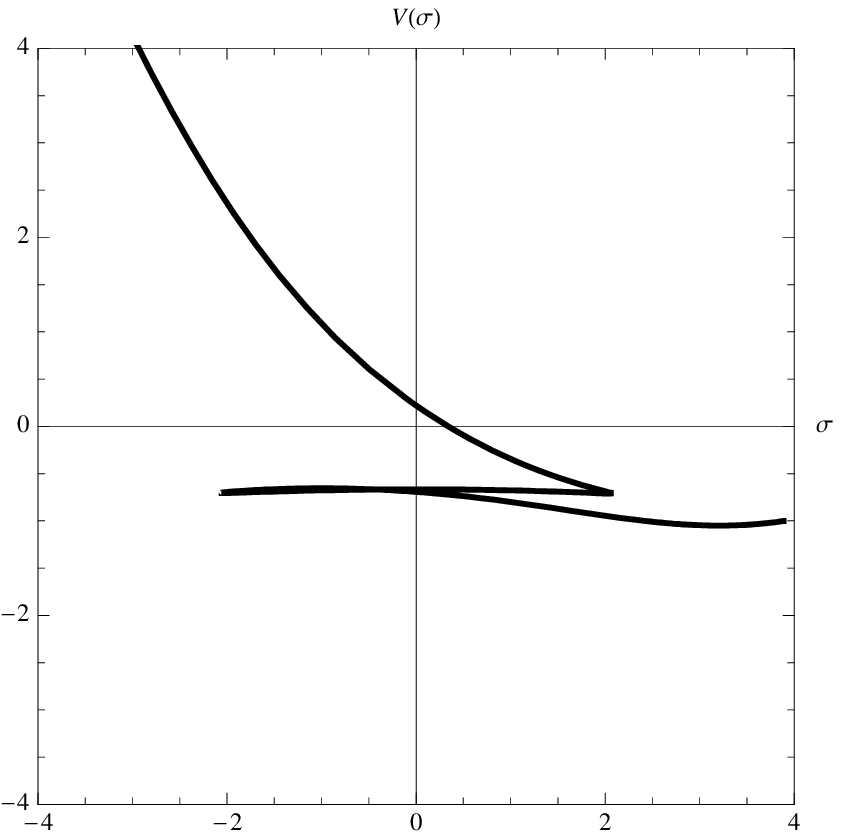}&
\includegraphics[scale=0.7]{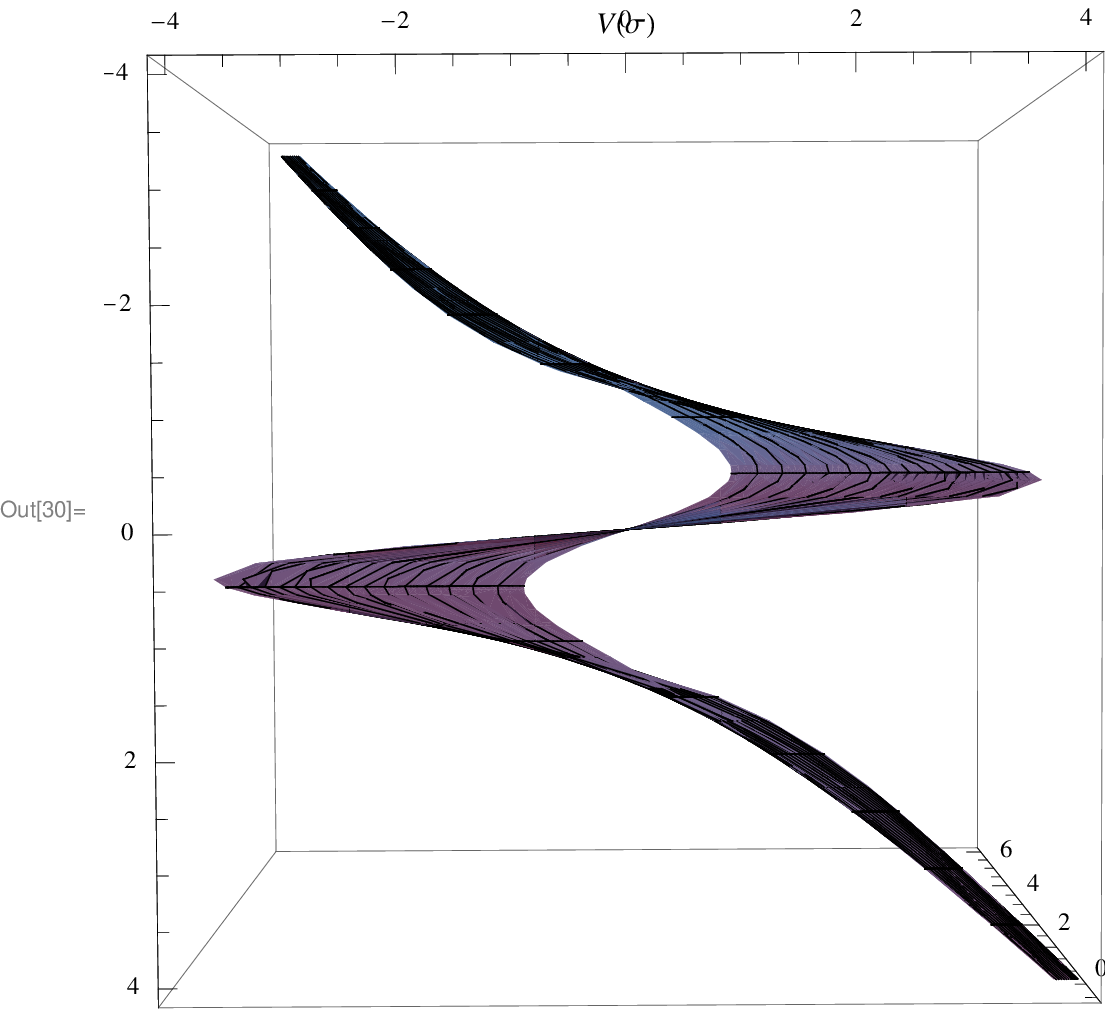}\tabularnewline
\hline
\end{tabular}
\caption { Parametric plot  of the  effective potential for $n=2$,
$\lambda=2$ and $\alpha=1/2$ in presence of $R^2$ correction in
two and three dimensions for the model (\ref{eq:HS}).}
\label{fig:3}
\end{figure}

For $n = 2, \lambda = 2$ and $\alpha = 0.5$, we have a large range
of the initial condition for which the scalar field evolves to the
minimum of the potential as shown in Fig. \ref{fig:3}. In conclusion, the introduction of $R^2$
term formally allows to avoid the singularity as it was suggested earlier
also in \cite{Abdalla, bamba, nojiriprd}.

\section{Conclusions}
\label{5} In this paper we have  discussed  the future finite-time
singularities which can, in principle, appear in dark energy
universe coming from modified gravity as well as in other dark
energy theories. Considering  $f(R)$ gravity models that satisfy
cosmological viability conditions (chameleon mechanism), it is
possible to show that finite-time singularities emerge in several
cases. Such singularities can be classified according to the
values of the scale factor $a(t)$, the density $\rho$ and the
pressure $p$ as done in Sect.\ref{2}. To avoid the singularities,
suitable boundary conditions have to be imposed which depend, in
general, from the parameters of the model as seen in Sect.\ref{3}.
It is interesting that in static spherically-symmetric spacetime
the finite-time singularity manifests itself as singularity at some
specific value of curvature.

Besides,  the problem can be analyzed by considering the mass of
an auxiliary scalar field coming from the further degrees of
freedom of $f(R)$-gravity. Such a scalar field is heavy in the
high-curvature regime whose density is much larger than the
present cosmological density. Such a field allows to study the
singularity problem using the conformal transformations. In this
case, we have to consider singularities of the scalar field and
the related effective potential and try to see if they can be
avoided in the conformal picture.  For example, the most striking
feature of the potential in Figure~\ref{fig:1}, and the core of
the problem for infrared-modified $f(R)$ models, is that curvature
singularity is at a finite distance  both in field and energy
values. The scalar field $\sigma$ directly feels the matter
distribution; for suitable values of the parameters, the force is
directed to the right, and drives the field $\sigma$ up the wall
toward infinite curvature. The characteristic scale of the
potential $V(\sigma)$ is the  curvature  $R_c$ which is of the
same order of magnitude of the today observed cosmological
constant. Such a value  is extremely low compared to the matter
densities we encounter at local Solar System and Galactic scales.
Given the scales involved, it is easy to drive the scalar field in
order to jump  the potential well considering the dynamics of
standard matter which could cause catastrophic curvature
singularity. If this were to happen, this is in contrast with any
viable model. Similarly, matter with sufficiently stiff equation
of state can destabilize the model by driving the field to
unstable points.
It is also remarkable that even in flat spacetime, some classes of the
classical potential may bring the theory to finite-time singularity.

Finally  we have discussed the possibility to address and solve
the singularity problem adopting higher-curvature corrections. In
such a way, the future cosmological era has no singularity as it
was demonstrated earlier in \cite{Abdalla,bamba,nojiriprd}. Then, it
also does not manifest itself in spherically-symmetric spacetime.
The interesting result of this analysis is the fact that
finite-time singularities, which present in some modified gravity, may
not influence the conjectured problem with relativistic star
formation indicated in ref.\cite{KobayashiMaeda}. Of course, this
fact depends on the specific characteristics of the adopted
modified gravity model which can be even totally free of future,
finite-time  singularities.


\section*{ACKNOWLEDGEMENTS}
This research is supported by INFN-CSIC bilateral project, by {\it
Azione Integrata Italia-Spagna 2007} (MIUR Prot. No. 464,
13/3/2006) grant and by MICINN (Spain) projects FIS2006-02842 and
PIE2007-50I023. The work by S.N. is supported by Min. of
Education, Science, Sports and Culture of Japan under grant no.
18549001.

\end{document}